\documentclass{article}

\usepackage{arxiv}

\usepackage[utf8]{inputenc} % allow utf-8 input
\usepackage[T1]{fontenc}    % use 8-bit T1 fonts
\usepackage{hyperref}       % hyperlinks
\usepackage{url}            % simple URL typesetting
\usepackage{booktabs}       % professional-quality tables
\usepackage{amsfonts}       % blackboard math symbols
\usepackage{nicefrac}       % compact symbols for 1/2, etc.
\usepackage{microtype}      % microtypography
\usepackage{lipsum}
\usepackage{graphicx}
\usepackage[bottom]{footmisc}
\usepackage[utf8]{inputenc}
\usepackage[T1]{fontenc}
\usepackage{multirow}
\usepackage{xr-hyper}
\usepackage[english]{babel}

\usepackage{hyperref, url, booktabs, amsfonts, nicefrac, microtype, lipsum, comment}

\usepackage{verbatim}
\usepackage{cite}
\usepackage{amsmath,amssymb,amsfonts}
\usepackage{algorithmic}
\usepackage{multicol, blindtext}
\usepackage{textcomp}
\usepackage[table,xcdraw]{xcolor}
\def\BibTeX{{\rm B\kern-.05em{\sc i\kern-.025em b}\kern-.08em
		T\kern-.1667em\lower.7ex\hbox{E}\kern-.125emX}}

\title{Supervised prediction of aging-related genes from a context-specific protein interaction subnetwork}

\author{
	Qi Li and Tijana Milenković *\\
	Department of Computer Science,  Center for Network \& Data Science, and Eck Institute for Global Health\\
	University of Notre Dame\\
	Notre Dame, IN 46556, USA \\
	\texttt{qli8@nd.edu, tmilenko@nd.edu} \\
}

\begin{document}
\maketitle

\begin{abstract}
\textbf{Background.} Human aging is linked to many prevalent diseases. The aging process is highly influenced by genetic factors. Hence, it is important to identify human aging-related genes. We focus on supervised prediction of such genes. Gene expression-based methods for this purpose study genes in isolation from each other. While protein-protein interaction (PPI) network-based methods for this purpose account for interactions between genes' protein products, current PPI network data are context-unspecific, spanning different biological conditions. Instead, here, we focus on an aging-specific subnetwork of the entire PPI network, obtained by integrating aging-specific gene expression data and PPI network data. The potential of such data integration \emph{has} been recognized but mostly in the context of cancer. So, we are the first to propose a supervised learning framework for predicting aging-related genes from an aging-specific PPI subnetwork. \\
\noindent\textbf{Results.} In a systematic and comprehensive evaluation, we  find that in many of the evaluation tests: (i) using an aging-specific subnetwork indeed yields more accurate aging-related gene predictions than using the entire network, and (ii) predictive methods from our framework that have not previously been used for supervised prediction of aging-related genes  outperform existing prominent methods for the same purpose. \\
\noindent\textbf{Conclusion.} These results justify the need for our framework.
\end{abstract}

% keywords can be removed

\section{Background \label{sect:intro}}

Human aging poses a risk for many prevalent complex diseases, such as diabetes, cancer, osteoarthritis, and cardiovascular or Alzheimer's disease \cite{campisi2013aging}. To better understand such diseases, it is important to study the aging process in human. This process is known to be highly influenced by genetic factors. Consequently, lots of effort has gone into identifying aging-related genes \cite{bolignano2014aging, fabris2017review}. Doing this via \emph{wet lab} experiments is hard because of ethical constraints and long human life span \cite{faisal2014dynamic}. So, \emph{computational} identification (i.e., prediction) of aging-related genes has received significant attention. This strategy uses features of genes extracted from some (e.g., gene expression or protein-protein interaction (PPI) data) data via unsupervised \cite{murabito2012search, yoo2017improving} or supervised \cite{freitas2011data, kerepesi2018prediction, fabris2017review} learning to predict genes that are most likely to be aging-related. Unsupervised and supervised learning differ in that the former does not consider current knowledge about aging when making predictions but instead considers it only when evaluating the predictions. The latter considers a part of the current knowledge when making predictions and its other part when evaluating the predictions. Both have advantages. Here, we focus on supervised learning, to benefit from the current knowledge while predicting new knowledge about aging.

There exits two major methodological directions for computationally predicting human aging-related genes: \emph{i}) from model species to human and \emph{ii}) within the human species. The former predicts a gene in human to be aging-related if its sequence \cite{tacutu2012human} or its PPI network neighborhood \cite{faisal2014global, gu2018homogeneous} is aligned to that of an aging-related gene in a model species. The latter predicts a gene in human to be aging-related if it is ``similar'' (see below) to known aging-related genes in \emph{human}. Both directions are important. They are simply different. Here, we focus on the latter. Methods of this type can be further divided into two major groups: gene expression-based ones \cite{jia2018analysis, lu2004gene, berchtold2008gene, holtman2015induction, simpson2011microarray} and PPI network-based ones \cite{fang2013classifying, fabris2016extensive, fabris2017review, farrell2016network, kerepesi2018prediction}, as follows. 

Gene expression-based methods predict a gene as aging-related if it is differentially expressed between younger age vs. older age, or if its expression value either increases or decreases with age when dealing with more than two ages \cite{jia2018analysis, lu2004gene, berchtold2008gene, holtman2015induction, simpson2011microarray}. A limitation of gene expression-based methods is that they study genes in isolation from each other. However, cellular processes, including aging, are carried out by genes' protein products interacting with each other in complex ways \cite{kirkwood2005understanding}. So, it is essential to consider interactions between proteins.

This is exactly what PPI network-based methods do. They predict a gene as aging-related according to how similar its position (i.e., node representation/embedding/feature) in the PPI network is to the network positions of known aging-related genes \cite{freitas2011data, fang2013classifying, fabris2016extensive, kerepesi2018prediction}. State-of-the-art approaches of this type are UniNet \cite{kerepesi2018prediction} and \emph{m}BPIs \cite{freitas2011data}. UniNet's feature concatenates 14 network centrality measures (e.g., degree or betweenness), where centrality of a node captures its importance in the network. The feature of \emph{m}BPIs is constructed as follows. First, $m$ nodes with the highest degrees in the network are identified. Then, for each node $v$ in the network, $v$'s feature has $m$ dimensions corresponding to the top $m$ highest-degree nodes, where each dimension $j$ of node $v$'s feature indicates whether $v$ interacts with the top $m$ highest-degree node $j$. A limitation of these and other PPI network-based methods is that the current PPI network data are \emph{context-unspecific}, meaning that the PPIs span different conditions (cell types, tissues, diseases, environments, patients, etc.) \cite{yeger2015human}; consequently, the current PPI network data are also \emph{static} \cite{faisal2014dynamic}. We argue that it is essential to consider aging-specific context of the entire PPI network, i.e., its \emph{aging-specific subnetwork}, by integrating aging-specific gene expression data with PPI network data. 

Indeed, the need for at least some level of data integration in the context of aging is being recognized. There exist approaches for supervised prediction of aging-related genes that extract genes' features from each of gene expression data and PPI network data. These two feature types are then integrated by concatenating them \cite{kerepesi2018prediction,freitas2011data}. However, methods of this type still extract features from each individual data type, i.e., they integrate the features rather than the data. Consequently, they still consider the entire (static) PPI network. 
 
Instead, we aim to first integrate the two data types and then extract node features from the resulting aging-specific subnetwork. The potential of such true data integration \emph{has} been recognized but in the context of \emph{cancer} \cite{wu2012integrating}. To the best of our knowledge, only two existing studies \cite{Elhesha2019,faisal2014dynamic}, including one by our own group \cite{faisal2014dynamic}, inferred a (dynamic) \emph{aging-specific} PPI subnetwork \cite{keshava2008human}. Specifically, given gene expression data for 37 ages \cite{berchtold2008gene} and an entire (static) PPI network \cite{keshava2008human}, in our past study \cite{faisal2014dynamic}, we formed 37 age-specific network snapshots, where each snapshot consists of all genes that are significantly expressed at the given age and all of their interactions. That is, a given age-specific network snapshot was formed by taking the induced subgraph on the significantly expressed genes at that age, which is why we refer to this aging-specific subnetwork inference approach as the \emph{induced approach}. Then, the 37 network snapshots were combined into a dynamic aging-specific PPI subnetwork, which captures how network positions of genes change with age. Note that the other existing study \cite{Elhesha2019} also used the induced approach to infer their dynamic, aging-specific network data.

The goal of our past study mentioned above \cite{faisal2014dynamic} was to \emph{infer} the \emph{dynamic} aging-specific subnetwork, because the aging process is clearly dynamic. That study \cite{faisal2014dynamic} did also predict aging-related genes, but in an \emph{unsupervised} manner, and primarily to validate the inferred aging-specific subnetwork. On the other hand, our goal here is to rely on exactly the aging-specific subnetwork from the existing study \cite{faisal2014dynamic} (that is, we do not deal with the problem of subnetwork inference) and to develop on top of it a comprehensive \emph{supervised} framework for predicting aging-related genes (and recall that unsupervised and supervised learning are different tasks). Also, the existing study \cite{faisal2014dynamic} predicted a gene as aging-related if its network position (namely, centrality) either significantly increased or decreased with age. As such, it would have missed any other types of changes of genes' network positions with age, such as network positions that continuously fluctuate or vary in a non-linear fashion. On the other hand, supervised prediction allows for learning network position change patterns of aging-related vs. non-aging-related genes automatically from the data.

\subsection{Our study and contributions}

\begin{figure}[h!]
    \includegraphics[width=0.95\linewidth]{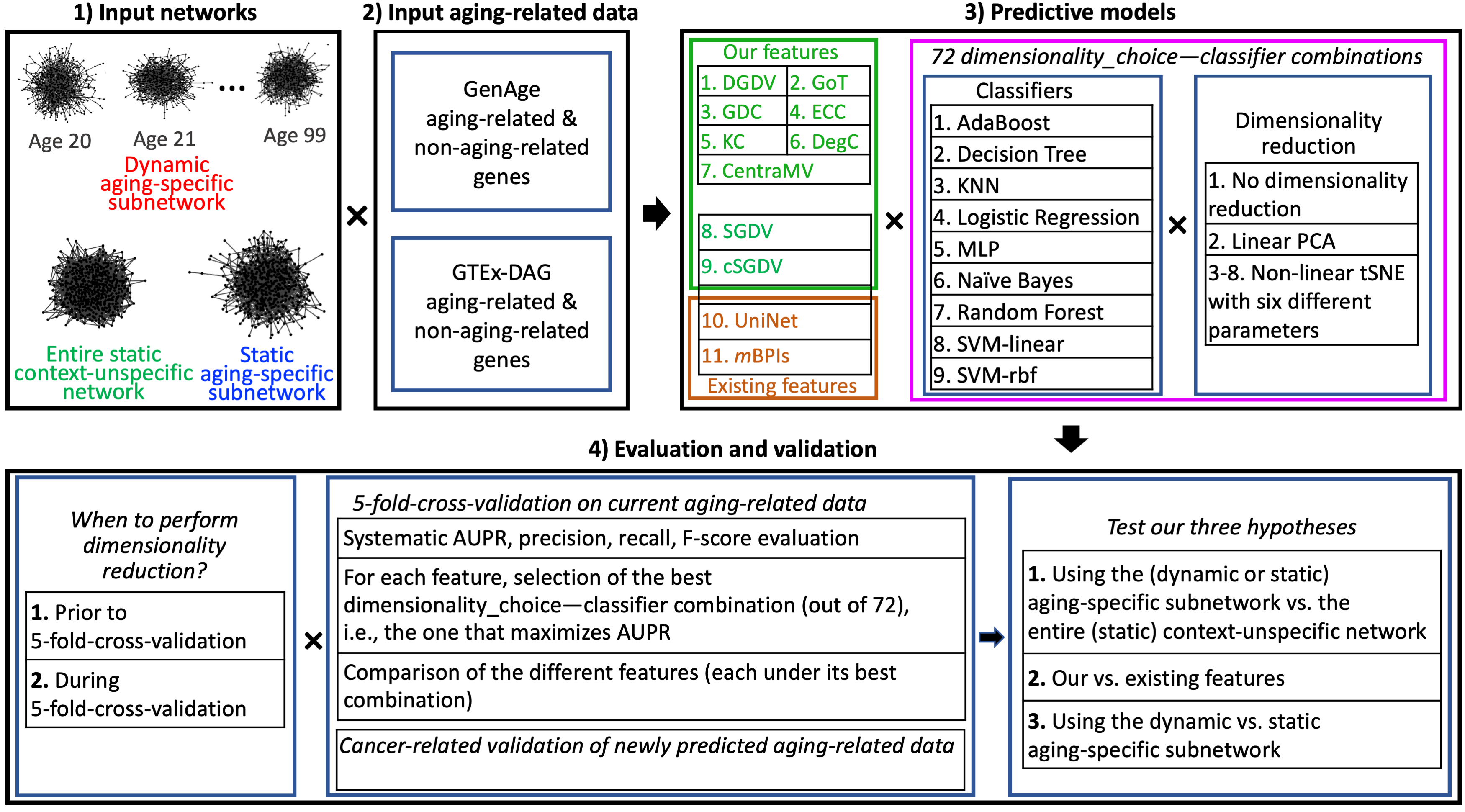}
  %  \vspace{-0.4cm}
    \caption{\textbf{Summary of AGENT and our study.} Note that features 1-7 work only on the dynamic aging-specific subnetwork. Features 8-11 work on the static aging-specific subnetwork as well as the entire static context-unspecific network. So, in total, there exist $7+4+4=15$ feature-network combinations. However, we could not run one of the features (cSGDV) on the entire context-unspecific network, as the large  size of this network makes this computationally prohibitive. So, in total, we have 14 feature-network combinations. For details on each step of the study, see the text.}
    \label{fig:workflow}
    \end{figure}

Unlike any of the existing studies, we present a supervised learning-based framework for predicting \textbf{ag}ing-related g\textbf{e}nes from an aging-specific PPI sub\textbf{n}e\textbf{t}work, AGENT (Figure \ref{fig:workflow}). 

%\textcolor{blue}{UPDATE FIG 1 WITH ALL FEATURES, DIM. RED METHODS/PARAMETERS, PRE-POST TRAIN, ETC.}

We test three hypotheses: (1) whether using an aging-specific subnetwork improves the supervised prediction accuracy compared to using the entire context-unspecific (static) network; (2) whether node features that are used in the task of supervised prediction of aging-related genes for the first time in this study outperform the existing UniNet and $m$BPIs features; and (3) because the aging process is dynamic, whether using the dynamic aging-specific subnetwork \cite{faisal2014dynamic} is superior to using its static version. \emph{To justify a need for our proposed AGENT framework, hypothesis 1 or hypothesis 2 must hold. Hypothesis 3 does not need to  hold; its outcome has no affect on our study's contributions.} In fact, finding that hypothesis 3, which by common sense is expected to be true, does not actually hold would be a more interesting contribution of our study than finding that it does hold. We test our hypotheses as follows. 

Because our goal here is not to propose a method for subnetwork inference, we rely on the existing aging-specific subnetwork, which is dynamic \cite{faisal2014dynamic}. Because we compare our proposed methods (see below) to the existing UniNet and $m$BPIs methods that can use only a static network, we also consider a static version of the dynamic subnetwork obtained by aggregating the PPIs from all age-specific snapshots of the dynamic subnetwork. We also consider the entire context-unspecific (also static) PPI network from which the dynamic and static aging-specific networks have been obtained. Thus, the three networks are as fairly comparable to each other as possible. 

Throughout our study, unless explicitly indicated otherwise, our primary definition of aging-related and non-aging-related genes is with respect to the GenAge database, a confident ground truth knowledge about human aging. (Later on, we  comment on using a secondary definition, i.e., a different aging-related data source, to see whether our findings are dependent on the choice of this parameter.) For the primary definition, of all 6,371 genes present in all three networks, we consider those present in GenAge \cite{tacutu2012human} as \emph{aging-related}. Also, of the 6,371 genes, we consider as \textbf{currently} \emph{non-aging-related} those genes that are not in GenAge and that also satisfy the following. We are aware of five other aging-related ground truth data \cite{jia2018analysis, lu2004gene, berchtold2008gene, simpson2011microarray}. These are not as confident as GenAge, which is why we do not consider their genes as aging-related in our primary definition. However, they likely contain some true aging-related knowledge. We want our non-aging-related genes to be as confident as possible, i.e., there should exist \emph{no} current ground truth evidence of their association to aging. So, we exclude all of the five data sets from what we consider as non-aging-related genes (in our primary definition). The above procedure results in 187 aging-related and 2,499 non-aging-related genes. It \emph{is} possible that some of the 2,499 genes are actually aging-related; it is just that no current evidence of this exists. So, we aim to prioritize for future experimental validation the 2,499 genes according to their likelihood of being aging-related. 

For each aging-related and non-aging-related gene, we use nine features that, while not entirely novel to our study, have not yet been used in this study's task of supervised prediction of aging-related genes. Among them, seven are \emph{dynamic}, i.e., extracted from the dynamic aging-specific subnetwork: dynamic graphlet degree vector (DGDV) \cite{hulovatyy2015exploring}, graphlet orbit transitions (GoT) \cite{aparicio2018graphlet}, graphlet degree centrality (GDC) \cite{milenkovic2011dominating}, eccentricity (ECC), $k$-core (KC), degree centrality (DegC), and centrality mean and variation (CentraMV) \cite{faisal2014dynamic}. The remaining two features are \emph{static}, i.e., extracted from the static aging-specific subnetwork and the entire (also static) context-unspecific PPI network: static graphlet degree vector (SGDV) \cite{milenkovic2008uncovering} and colored (or heterogeneous) SGDV (cSGDV) \cite{gu2018homogeneous}. SGDV is DGDV's fairly comparable static counterpart. Both of these are homogeneous. In contrast, cSGDV is SGDV's fairly comparable heterogeneous counterpart. Of the nine features, DGDV, GoT, GDC, SGDV, and cSGDV are graphlet-based; graphlets are  subgraphs (Lego-like building blocks) of a network. We use these features because graphlets have been shown to be the state-of-the-art in many tasks \cite{newaz20195} (but not yet in this study's task). GDC, ECC, KC, DegC, and CentraMV are centrality-based. These are dynamic counterparts of UniNet, against which we compare our nine features; also, we compare against $m$BPIs. Just as the (dynamic or static) graphlet-based features, the dynamic centrality-based features have not yet been used in this study's task. 

We couple each of our nine and the two existing features with nine prominent machine learning classifiers to make aging-related gene predictions. For each feature, prior to using it in a classifier, we examine whether reducing its dimension via a linear or non-linear technique yields improvement (in terms of protein function prediction accuracy -- see below) compared to using the full feature; we examine feature dimensionality reduction as it could remove potential redundancy between the different (often many) dimensions. Then, for each feature, we aim to choose whichever dimensional reduction technique (including no reduction) and classifier yield the most accurate predictions. We evaluate each feature--dimensionality\_choice--classifier combination via 5-fold cross-validation: we train on a subset of our aging- and non-aging-related genes and test prediction accuracy on the remaining aging- and non-aging-related genes.  We evaluate prediction accuracy in terms of the area under the precision-recall curve (AUPR) over all prediction thresholds, as well as via precision, recall, and F-score at the prediction threshold where F-score is maximized. Then, we compare the 11 considered features (each under its best dimensionality choice and classifier) in terms of their prediction accuracy (as discussed above). Also, we compare them by measuring how many of their predictions that are not in any of the considered aging-related ground truth data we can validate in cancer-related data, as cancer is known to be highly related to aging \cite{campisi2013aging}. In the context of the cancer-related validation of the newly predicted aging-related genes, we again measure precision, recall, and F-score (AUPR is not applicable to this analysis).

To validate our hypothesis 1, it would suffice for the best feature on dynamic or static aging-specific subnetwork to outperform every feature on the entire static context-unspecific network. Indeed, this is what we find in the 5-fold cross-validation for precision, recall, and F-score, although not for AUPR. Also, this is what we find in the validation on cancer data for all of precision, recall, and F-score. So, hypothesis 1 holds in almost all evaluation/validation tests.

To validate our hypothesis 2, it would suffice for one or more of our nine features on the dynamic or static aging-specific subnetwork (depending on the feature) to be superior to both of the existing features on the static aging-specific subnetwork (as these two features can be run on only the static but not dynamic subnetwork). Indeed, this is what we find in both the 5-fold cross-validation and the cancer-related validation for precision, although not for recall, F-score, and AUPR.

Our hypothesis 3 was already confirmed in an \emph{unsupervised} analysis of the same aging-specific subnetworks \cite{hulovatyy2015exploring}. To validate  it in our \emph{supervised} analysis, it would suffice for the best feature on the dynamic subnetwork to be superior to the best feature on the static subnetwork. Surprisingly, we find hypotheis 3 to hold only in the 5-fold cross-validation and only for precision. That is, it does not hold with respect to any other measure in the 5-fold cross-validation, nor with respect to any measure in the validation on cancer data. So, hypothesis 3 holds in a single evaluation/validation test.
Implications of this unexpected result are discussed in Section \ref{sect:results_dynamic_static}. Importantly, we again note that the outcome of hypothesis 3  has  no  affect  on  our  study’s contributions. The other two hypotheses (1 and 2), at least one of which \emph{must} hold, \emph{have} both been shown to hold overall. 

The above results are when using GenAge to define aging- and non-aging-related genes. Human genes in GenAge have been implicated in aging mostly because they are sequence-based orthologs of expertimentally validated aging-related genes from model species. However, the human species has unique biological aspects compared to model species. So, using the GenAge data to train a classifier and then make  aging-related gene predictions in human might miss the human-unique aspects \cite{danchin2018bacteria}.  Hence, it would be interesting to define aging- and non-aging-related genes with respect to data obtained directly by studying the human species . The only such data comes from analyzing expression levels of human genes at different ages. We use such data (from the genotype-tissue expression  (GTEx) project \cite{jia2018analysis}) to define an alternative set of aging- and non-aging-related genes. Shockingly, we find that it is hard to make sense of the predictions resulting from using the GTex data to train a classifier and then make aging-related gene predictions. For example, while for the GenAge data, the classification accuracy of the predictions is in the 60\%-70\% range for almost all approaches and all accuracy measures, for the GTex data, the classification accuracy is almost always under 30\%. We comment more on these results in Section \ref{sect:secondaryGTEx-result}.

\vspace{-0.05cm}
\section{Methods}

\subsection{Data}
\subsubsection{The three considered networks}
Table \ref{table:net-size} shows sizes of: (\emph{i}) the dynamic aging-specific PPI subnetwork, (\emph{ii}) its corresponding static aging-specific PPI subnetwork, and (\emph{iii}) the entire (also static) context-unspecific PPI network.  The latter is the network used to infer the two aging-specific subnetworks in the first place \cite{faisal2014dynamic}, namely the PPI network from HPRD \cite{keshava2008human}. This allows for a fair comparison of the three networks. Using a different (e.g., more recent) entire context-unspecific PPI network would make it hard to fairly evaluate our hypotheses without having to infer the new corresponding aging-specific subnetworks. Not only is this out of the scope of our study, but also, we \emph{have} compared the entire context-unspecific HPRD PPI network against a more recent (2019) entire context-unspecific BioGRID PPI network \cite{stark2006biogrid}. We have found that for all considered static features (the only features that can be run on the two static networks), performance is better on the HPRD network than on the BioGRID network (results not shown). Throughout our study, we aim to give each feature the best-case advantage. Clearly, the HPRD network is more advantageous. Thus, we do not further consider the BioGRID network.

\begin{table}[!htp]
\centering
\caption{Sizes of the considered networks. We express the size for the dynamic subnetwork as  the average over its 37 snapshots.}
\vspace{-0.05in}
\label{table:net-size}
\begin{tabular}{@{}l|c|c@{}}
\hline
\ Network &\# of nodes&\# of edges \\ \hline
\ Dynamic aging-specific subnetwork & 4,696 & 15,062  \\ 
\ Static aging-specific subnetwork & 6,371 & 22,975  \\ 
\ Entire static context-unspecific network & 8,938 & 35,900\\ \hline
\end{tabular}
%\vspace{-0.05in}
\end{table}

\subsubsection{The six considered aging-related ground truth data sets} \label{subsect:sixagingdata}

\begin{itemize}
    \item Tacutu \textit{et al.} \cite{tacutu2017human} identified 307 human aging-related genes, mostly sequence orthologs of aging-related genes in model organisms. Of these, 187 are present in all three networks and we denote them as \underline{GenAge}; this is the name of the trustworthy database they come from. 

    \item Jia \textit{et al.} \cite{jia2018analysis} identified \textbf{two} sets of aging-related genes by studying the GTEx data. One set includes 710 genes whose expressions are \emph{upregulated} with age. Of these, 239 are present in all three networks and we denote them as \underline{GTEx-UAG}. The other set includes 863 genes whose expressions are \emph{downregulated} with age. Of these, 439 are present in all three networks and we denote them as \underline{GTEx-DAG}.

    \item Lu \textit{et al.} \cite{lu2004gene} identified 442 genes whose expressions in the brain  change with age. Of these, 303 are present in all three networks and we denote them as \underline{BEx2004}.

    \item Berchtold \textit{et al.} \cite{berchtold2008gene} identified 8,277 genes whose expressions in the brain  change with age. Of these, 2,918 are in all three networks and we denote them as \underline{BEx2008}. 

    \item Simpson \textit{et al.} \cite{simpson2011microarray} identified 2,911 genes whose expressions in the brain  change across different stages of Alzheimer's disease. Of these, 1,162 are present in all three networks and we denote them as \underline{ADEx2011}. 
\end{itemize}

\subsubsection{The primary definition of aging-related genes and non-aging-related genes with respect to GenAge \label{sect:primaryGenAge}}
Of the 6,371 genes present in all three networks, we treat the 187 GenAge genes as \underline{our aging-related genes}. Of the 6,371 genes, we treat those 2,499 genes that are not in any of the six ground truth data sets as \underline{our non-aging-related genes}.

\subsubsection{The secondary definition of aging-related genes and non-aging-related genes with respect to GTEx-DAG \label{sect:secondaryGTEx}}

In Section \ref{sect:secondaryGTEx-result}, we test whether our findings change if we change the choice of aging-related data used to define aging- vs. non-aging-related genes. So,  instead of the primary definition of the two gene sets from Section \ref{sect:primaryGenAge} that uses the data from GenAge, in Section \ref{sect:secondaryGTEx-result}, we use a secondary definition of aging- vs. non-aging-related genes that relies on the data from GTEx, as follows.

GTEx consists of GTEx-DAG and GTEx-UAG. The two gene sets have very different properties. That is, GTEx-DAG genes  are more evolutionary conserved,  are more enriched in essential genes, are critical for PPIs, and do not show tissue-specific gene expression patterns, while GTEx-UAG genes are less evolutionary conserved, are less enriched in essential genes, are not critical for PPIs, and show tissue-specific gene expression patterns. Given that GTEx-DAG genes are critical for PPIs while GTEx-UAG genes are not, and given that we predict aging-related genes from PPI network data, we expect GTEx-DAG to be more relevant for our predictive task than GTEx-UAG. So, we define the secondary set of aging- vs. non-aging-related genes with respect to GTEx-DAG.

Specifically, in the secondary definition, of the 6,371 genes present in all three networks, we treat the 439 GTEx-DAG genes as the aging-related genes. Of the 6,371 genes, we treat those 2,499 genes that are not in any of the six ground truth data sets as the non-aging-related genes. 

Importantly, we use the secondary GTEx-DAG-based definition only in Section \ref{sect:secondaryGTEx-result}. In all other parts of the paper, we use the primary GenAge-based definition from Section \ref{sect:primaryGenAge}.

\vspace{0.1cm}

\subsubsection{Cancer-related genes \label{subsect:cancerdata}}

%\textcolor{green}{FIGURE OUT WHETHER YOU WILL USE THE DRIVER GENES OR NOT, AND IF NOT, UPDATE THIS SECTION}

\begin{itemize}
\vspace{-0.05cm}
    \item Vogelstein \textit{et al.} \cite{vogelstein2013cancer} identified 138 mutation-related cancer driver genes, i.e., genes whose intragenic mutations contribute to cancer. Of these, 100 genes are present in all three networks and we denote them as \underline{mutation driver genes}. Of the 100 genes, 22 are aging-related and 29 are non-aging-related according to the primary GenAge-based definition  from Section \ref{sect:primaryGenAge}. Of the 100 genes, 6  are aging-related and 29 are non-aging-related according to the secondary GTex-DAG-based definition from Section \ref{sect:secondaryGTEx}.
    
    \item Sondka \textit{et al.} \cite{sondka2018cosmic} identified 723 cancer driver genes by manually curating cancer-related genes from COSMIC to determine the presence of somatic mutation patterns in cancer. Of these, 469 genes are present in all three networks. Combining the 469 genes with the 100 mutation driver genes, we obtain 474 genes, which we denote  as \underline{all driver genes}. Note that the mutation driver genes are a subset of the all driver genes. Of the 474 genes, 53 are aging-related and 135 are non-aging-related according to the primary GenAge-based definition  from Section \ref{sect:primaryGenAge}. Of the 474 genes, 21  are aging-related and 135 are non-aging-related according to the secondary GTex-DAG-based definition from Section \ref{sect:secondaryGTEx}.
\end{itemize}

\subsection{Node features and classifiers}

\subsubsection{The 11 considered node features}

\begin{itemize}
\vspace{-0.05cm}
\item \underline{DGDV} of a node \cite{hulovatyy2015exploring} counts, for each dynamic graphlet on up to $n$ nodes and $e$ events (temporal edges), how many times the node  participates in (touches) the given dynamic graphlet's topologically unique node positions called automorphism orbits. Dynamic graphlets are an extension of static graphlets, i.e., up to $n$-node subgraphs of a static network, to the dynamic setting, with temporal information added onto edges of static graphlets; this information indicates the temporal order in which events occur in a dynamic network. When we compute DGDV, we consider up to 4-node and 6-event dynamic graphlets (which have 3,727 orbits), both because this was suggested in the DGDV publication \cite{hulovatyy2015exploring}, as well as because considering larger dynamic graphlets would be computationally prohibitive on our data.

\item \underline{GoT} \cite{aparicio2018graphlet} is another dynamic extension of static graphlets. GoT of a node counts how many times the node's participation in one graphlet (e.g., a triangle) changes to its participation in another graphlet (e.g., a 3-node path, when an edge of a triangle ``breaks'') between every two consecutive time points, and for every pair of considered graphlets (i.e., their orbits). When we compute GoT, we consider 4-node graphlets, as done in the GoT publication \cite{aparicio2018graphlet}, and to make GoT fairly comparable to DGDV. This results in examining 11 orbits and thus 121 orbit pairs. 

\item \underline{GDC} of a node \cite{milenkovic2011dominating} is a weighted sum of the counts of all graphlets on up to $n$ nodes (i.e., their orbits) in which the node participates, where the weights account for orbit dependencies (see the GDC publication \cite{milenkovic2011dominating} for details). According to GDC, the more   (larger and denser) graphlets a node touches, the more complex its extended network neighborhood, and so the more central it is. 

\hspace{0.4cm}This definition of GDC is for a static network. We use such GDC to compute, for each node, its centrality in each of the 37 snapshots of the dynamic subnetwork. Then, we combine a given node's 37 GDC values into its 37-dimensional dynamic GDC feature. We do the same for the other centrality features below (ECC, KC, and DegC), each resulting in a 37-dimensional node feature.

\hspace{0.4cm}While for DGDV and GoT we consider up to 4-node graphlets (for reasons discussed above), for GDC, we have to use up to 5-node graphlets. This is because the GDC code that we used, which originates from \cite{faisal2014dynamic}, is an executable only that is a part of a larger software. The latter does not allow for specifying the desired graphlet size; instead, it uses up to 5-node graphlets by default.  While in this way, GDC uses more of network topology than the features that use up to 4-node graphlets, as we show in Section \ref{sect:results}, GDC is \emph{not} the best graphlet feature. 

\item \underline{ECC} of a node \cite{faisal2014dynamic} is the reciprocal of the shortest path distance from the node in question to the farthest of all other nodes in the network. 

\item \underline{KC} of a node \cite{faisal2014dynamic} is based on the notion of a network core. A core of a network is a subgraph in which each node is connected to at least $k$ other nodes in the subgraph. So, a network has its 1-core, its 2-core, its 3-core, etc. KC of a node is $k$ if the node is in the $k$-core.

\item \underline{DegC} of a node \cite{faisal2014dynamic} is the number of edges it touches. 

\item \underline{CentraMV}  \cite{faisal2014dynamic} works as follows. For a given centrality-based feature (out of  GDC, ECC, KC, and DegC), the mean and the corresponding variation are computed over a given node's 37 centrality values corresponding to the 37 snapshots of the dynamic subnetwork. The mean is self-explanatory, and the variation of node $u$ is $var(u) = \sum_{i= 1}^{36}(centrality(u)_{i+1} - centrality(u)_i)/36$, $i = 1, 2, ..., 36$. These two quantities are computed for each of the four centrality-based features, and the resulting eight values form the CentraMV node feature.

\item \underline{SGDV} of a node \cite{milenkovic2008uncovering} counts, for each static graphlet on up to $n$ nodes, how many times the node participates in the given graphlet (i.e., in each of its orbits). When we compute SGDV \cite{hovcevar2014combinatorial}, for a fair comparison with DGDV, we use up to 4-node graphlets, which encompass 15 orbits. Additionally, for a fair comparison with GDC, we also test SGDV when using up to 5-node graphlets, which encompass 73 orbits. While SGDV based on up to 5-node graphlets uses more of network topology than SGDV based on up to 4-node graphlets, we find that the former does \emph{not} improve upon the latter (results not shown). Consequently, for simplicity, in the paper, we only focus on SGDV based on up to 4-node graphlets.

\item \underline{cSGDV} of a node \cite{gu2018homogeneous} counts, for each node-colored (i.e., heterogeneous) graphlet on up to $n$ nodes and with $c$ colors, how many times the node participates in the given node-colored graphlet, i.e., in each of its orbits. Node-colored graphlets are an extension of regular graphlets (i.e., static and homogeneous ones, as used in SGDV above)  to the (still static but now) heterogeneous setting, with heterogenous node information added as colors onto nodes of homogeneous graphlets. Given all up to $n$-node homogeneous graphlets with $o$ orbits, and  given $c$ node colors, there are $o \times (2^c - 1)$ node-colored graphlet orbits, i.e., the length of a node's cSGDV is $o \times (2^c - 1)$. In this study, we use aging-related information about genes as heterogenous information, i.e., node colors. We consider $c=4$ colors, as follows. When computing a node $v$'s cGDV, we do not want to leak the information of whether node $v$  is  aging-related or not to its feature, as this could cause a circular argument during classification. So, we ignore such information entirely by assigning a ``neutral'' color, referred to as color 1, only to node $v$ out of all nodes in the network when computing $v$'s cGDV. At the same time, we label each of the other nodes in the network with one of three other colors, depending on which of the following three gene sets it belongs to according to Section \ref{sect:primaryGenAge}: (i) the aging-related 187-gene set (color 2), (ii) the non-aging-related 2,499-gene set (color 3), or (iii) neither the aging- or non-aging-related gene set (color 4). We repeat the above steps for each node $v$ that we use in classification. This way, for heterogeneous graphlets with four node colors: (i) when considering up to 4-node graphlets, which have 15 orbits, cGDV of a node encompasses $15 \times (2^4-1)=255$ orbits; (ii) when considering up to 5-node graphlets, which have 73 orbits, cGDV of a node encompasses $73 \times (2^4-1)=1,095$ orbits. Just as SGDV, we also test cSGDV on graphlets with up to four nodes as well as cSGDV on graphlets with up to five nodes (both with the same four colors).  We again find that the latter does \emph{not} improve upon the former (results not shown). So, again, for simplicity, in the paper, we only focus on cSGDV based on up to 4-node graphlets.

\item \underline{UniNet}'s feature \cite{kerepesi2018prediction} combines 14 node centralities: DegC, ECC, KC, average shortest path, betweenness, closeness, clustering coefficient, neighborhood connectivity, radiality, stress, topological coefficient, aging neighbor count, aging neighbor ratio, and binary aging neighbor representations (for details, see the UniNet publication \cite{kerepesi2018prediction}). 

\item The feature of \underline{$m$BPIs} \cite{freitas2011data} is defined in Section \ref{sect:intro}. We have evaluated three $m$ values, 10, 20, and 30, as suggested in the $m$BPIs paper \cite{freitas2011data}.  Because in our preliminary analyses we have found that $m=30$ performs the best, we perform the full analyses (described below) and thus report results only  for this value of $m$. Consequently, henceforth, we refer to $m$BPIs as 30BPIs.
\end{itemize}

\subsubsection{Choice of feature dimensionality: no reduction vs. (non)linear reduction} \label{subsect:featuredimension}

Some of the considered features, especially DGDV and GoT, have high dimensions. High dimensionality of a feature can lead to its overfitting during classification \cite{everitt2002cambridge}. Because of this, and as typically done \cite{hulovatyy2015exploring,aparicio2019temporal,gu2018homogeneous}, for each feature,  we explore the effect of reducing its dimensionality via a prominent linear as well as non-linear dimensionality reduction technique, as follows.

First, we consider no dimensionality reduction.

Second, we apply the \emph{linear} principal component analysis (PCA) to reduce the feature dimension, while considering as few PCA components as needed to account for at least 90\% of variation.

Third, we apply the \emph{nonlinear} t-distributed stochastic neighbor embedding (tSNE) \cite{maaten2008visualizing} that reduces the feature dimensionality into a 2-dimensional space. In the process, we test six  perplexity values  within the $[5,50]$ range suggested in tSNE's original publication \cite{maaten2008visualizing}. Specifically, we test these values: 5, 13, 21, 30, 40, and 50. Perplexity indicates the  number of nearest neighbors of each node in the data that balances the attention between local and global aspects of the data. In other words, larger perplexity corresponds to tSNE treating the data as having a higher density and smaller perplexity corresponds to tSNE treating the data as having a lower density.

So, in total, we consider $1+1+6=8$ dimensionality choices.

%\medskip

\subsubsection{The nine considered classifiers}

Adaptive Boosting (\underline{AB}), Random Forest (\underline{RF}), and Decision Tree (\underline{Dtree})  combine a set of trees to improve classification performance. Logistic Regression (\underline{LR})  uses a logistic function to model a binary dependent variable. Naïve Bayes (\underline{NB}) is a family of ``probabilistic classifiers'' that apply Bayes’ theorem assuming that the features are independent. We use Gaussian NB, which extends the traditional NB to handle real-valued features. $K$-Nearest Neighbors (\underline{KNN})  is a non-parametric method that assigns an object to the class that is the most common among its $K$ nearest neighbors. Support Vector Machine (SVM) outputs an optimal hyperplane by using the hinge loss function to categorize objects. It can be used with multiple kernel functions based on whether the objects are linearly separable. We use both a linear (\underline{SVM-linear}) and non-linear kernel (radial basis function (\underline{SVM-rbf})). Multilayer Perceptron (\underline{MLP})  is a class of feedforward artificial neural network that can classify objects when they are not linearly separable.

We implement the classifiers in scikit-learn (version 0.20.3) \cite{scikit-learn}, a public machine learning library for Python. We use the classifier's default parameter values. For some parameters (e.g., $K$ in KNN), we \emph{have} tested several values around the default value. The default value (i.e., $K = 5$ in KNN) \emph{is} the best. 

\vspace{-0.1cm}
\subsection{Evaluation and validation \label{sect:methods_evaluation} }

All of the following methodology is described with respect to the primary GenAge-based definition of aging- and non-aging-related genes. The methodology for the secondary GTEx-DAG-based definition is analogous.

\vspace{-0.4cm}

\subsubsection{Evaluation via 5-fold cross-validation,  prediction accuracy measures, and selection of the best dimensionality choice and classifier for each feature} \label{subsect:5-fold-cv}

For each of the 187 aging- and 2,499 non-aging-related genes, for each of the 11 features, for each feature's eight dimensionality choices (no reduction, PCA, and six tSNE versions),  we compute the given gene's feature from the corresponding network. Then, for each of the 11 features, for each of its eight dimensionality choices, we train and test each of the nine classifiers using 5-fold cross-validation.

Namely, for a given classifier, we randomly divide each of the aging- and non-aging-related gene sets into five equal-sized subsets. Then, we train a classifier on the union of four aging- and four non-aging-related gene subsets and test on the union of one aging- and one non-aging-related subset. We repeat this five times so that each time we are using a different subset as the testing data. The output is a probability of each gene being aging-related. We predict the top $g$ most probable genes as aging-related, where we vary $g$ from 1 to 60 in increments of 2 and then from 60 to 500 in increments of 10. 500 is the approximate size of a testing data set (i.e., of a predicted set) in each fold.

Given a prediction set and the set of our aging-related genes (GenAge), a true positive (TP) is a gene that is in GenAge and is predicted as aging-related, a false positive (FP) is a gene that is not in GenAge but is predicted as aging-related, and a false negative (FN) is a gene that is in GenAge but is not predicted as aging-related. In our data, there are more non-aging- than aging-related genes. So, we use the above quantities to calculate three popular prediction accuracy measures that \emph{can} deal with such imbalanced data: \underline{precision} (the percentage of the predicted aging-related genes that are among the GenAge aging-related genes), \underline{recall} (the percentage of the GenAge aging-related genes that are among the predicted aging-related genes), and \underline{F-score} (the harmonic mean of precision and recall), each averaged over the five runs of the cross-validation. 

We compute these measures for each prediction set, i.e., at each value of $g$. We summarize the performance of a  feature--dimensionality\_choice--classifier combination over the entire $g$ range via the \underline{area under the precision-recall curve (AUPR)}. For each feature, we choose the dimensionality\_choice--classifier combination that maximizes the AUPR.

Then, we compare the different features' AUPRs. In addition, we compare the features as follows. AUPR does not produce a specific predicted gene set for further analysis. To produce such a set, we need to choose a specific value of $g$. Choosing a best-case value of $g$ is non-trivial, as there is a trade-off between precision and recall. We believe that a meaningful value of $g$ is the one where F-score (which balances between precision and recall) is maximized. So, for each feature (under the dimensionality\_choice--classifier combination that maximizes its AUPR), we choose such a $g$ value. Then, we compare the different features in terms of precision, recall, and F-score at the selected $g$ value.

\underline{We favor precision over recall (and thus over F-score as well as AUPR)} because we believe that in biomedicine, for wet lab validation of predictions, it is more important to have say 90 correct predictions out of 100 made than say 300 correct predictions out of 1,000 made. While in the latter case many more are correct (300 vs. 90), which would likely yield a higher (approximately triple) recall, in the latter case also many more are incorrect (700 vs. only 10), which results in a lower precision (0.3 vs. 0.9). While we favor higher precision, at the same time, recall should not be too low, which is why we also consider recall as well as F-score, i.e., why we choose $g$ where F-score is maximized.

\subsubsection{Choice of when to reduce feature dimensionality: prior to vs. during the 5-fold cross-validation \label{subsect:priororduring}} 

A feature's dimensionality can be reduced in two established ways. First, this can be done prior to the 5-fold cross-validation. This means that the feature dimensionality is first performed for the whole data set. Then, the dimensionality-reduced features are split into the training and testing portions of the data and used in the 5-fold cross-validation. Second, this can be done during the 5-fold cross-validation. This means that the raw data (without any dimensional reduction yet) is first split into the training and testing portions. Then, for each fold of the cross-validation, the following three steps are performed. 1) A given dimensionality reduction method is applied only to the training data set. 2) The dimensionality-reduced training data  set is used to train a model (i.e., classifier). 3) Given the testing data set, the trained model is used to reduce the dimensionality of the testing data set into the same space as the training data set, and the dimensionality-reduced testing dataset is then used to evaluate the prediction accuracy for the model. 

Both the feature dimensionality reduction prior to the cross-validation \cite{wood2007classification, zhu2008selection, newaz2018network} and the feature dimensionality reduction during the cross-validation \cite{fortuna2004improved, hastie2009elements, boulesteix2009optimal} the cross-validation are widely used in the supervised prediction literature. The principle of supervised prediction requires ``perfect separation'' such that none of the data from the testing data set should be leaked to the training data set while training the predictive model \cite{daumer2008reducing}. The feature dimensionality reduction prior to the cross-validation violates this principle, while the feature dimensionality reduction during the cross-validation does not. This is why throughout this paper, we mainly focus on the latter. However, the choice between the two options is still debatable, as it depends on many factors, including but not limited to data sample size \cite{hornung2015measure}. So, we also consider the former, but we do so only in Section \ref{sect:pretrainreduction}.

\subsubsection{Evaluation: statistical significance of results}
  
We compare the prediction accuracy of each feature (under its best dimensionality \\
\_choice-classifier combination) with that of a random approach, which works as follows. For a given testing data set and a given $g$ value, we randomly select $g$ genes from the testing data and predict them as aging-related. Then, we calculate prediction accuracy (precision, recall, and F-score). We do this for the testing data in each of the five folds. We repeat the above procedure 30 times, to account for its randomness. This results in $5 \times 30 = 150$ random runs. For the random approach, we report its prediction accuracy averaged over the 150 runs.

For a prediction accuracy measure, given two features (each under its best dimensionality\_choice--classifier combination), i.e., given each feature's five prediction accuracy values resulting from the five folds, we compute the statistical significance (i.e., $p$-value) of the difference in the two features' prediction accuracy via the paired Wilcoxon rank sum test. Since we run this test  multiple times, we apply false discovery rate (FDR) correction to adjust the $p$-values.

\vspace{0.1cm}

\subsubsection{Validation of newly predicted genes on cancer data}
We denote as \underline{newly predicted} those genes that are predicted as aging-related (by a given feature under its best dimensionality\_choice--classifier combination) but are not present in any of the six aging-related ground truth data sets. Because aging and cancer are known to be tightly linked \cite{campisi2013aging}, the higher the overlap between the newly predicted genes and the genes from the cancer-related data (i.e., either the 188 all driver genes or the 51 mutation driver genes), the more accurate the predictions. For each of the all driver genes and mutation driver genes, we quantify the overlap size via \underline{precision} (the percentage of the newly predicted genes that are among the cancer genes), \underline{recall} (the percentage of the cancer genes that are among the newly predicted genes), and \underline{F-score} (the harmonic mean of precision and recall).

\vspace{0.1cm}

\subsubsection{Overlap between prediction sets} We use the Jaccard index (J) \cite{vijaymeena2016survey} to measure the size of the overlap between two prediction sets $A$ and $B$: $\frac{|A \cap B|}{|A \cup B|}$. The lower its value, the more complementary sets $A$ and $B$ are. We evaluate the statistical significance (i.e., $p$-value) of the given overlap size via the hypergeometric test. Since we run this test multiple times, we apply FDR correction to adjust the $p$-values.

%%%%%%%%%%%%%%%%%%%%%%%%%%%%%%%%%%%%%%%%%%%%%%%%%%%%%%%%%%%%%%%%%%%%%%%%%%%%%%%%%%%%%%%%%%%%%%%%%%%%%%%%%%%%%%%%%%%%%%%%%%%%%%%%%%%%%%%%%%%%%%%%%%%%%%%%%%%%%%%%%%%%%%%%%%%%%%

\section{Results and discussion \label{sect:results}}

All results except those in Section \ref{sect:secondaryGTEx-result} are for the primary GenAge-based definition of aging- vs. non-aging-related genes. 
Also, all results except those in Section \ref{sect:pretrainreduction} are for the case when feature dimensionality reduction is done during the 5-fold cross-validation.
\vspace{-0.1cm}
\subsection{Selecting the best dimensionality choice and classifier combination for each feature}

Recall from Figure \ref{fig:workflow} that of all 11 considered features, seven work on the dynamic subnetwork, and four work  on the static subnetwork as well as the entire static network. So, in total, we have $7+4+4=15$ feature-network combinations. However,  we could not run one of the features (cSGDV) on the entire context-unspecific network, as the large  size of this network makes this computationally prohibitive. So, in total, we have 14 feature-network combinations. Henceforth, for simplicity, we just say that we have 14 features, though again we note this includes some cases where the same feature is run on different networks.

Recall that for each of the 14 features, we choose the best one of the 72 combinations of eight dimensionality choices (no reduction, PCA, and six tSNE versions) and nine classifiers. By ``best'', we mean the one that maximizes AUPR. The selected choices for each feature are shown in Table \ref{table:classifiers}. Briefly, in terms of the dimensionality choice, for seven of the 14 features, no reduction wins over PCA and tSNE, while for the remaining seven, PCA wins over no reduction and tSNE. In other words, tSNE never wins. In terms of the classifier choice, Logistic LR is overall the best classifier, which wins for seven out of the 14 features, and it is followed by four other classifiers that win for one to three of the remaining seven features. %\textcolor{blue}{COMMENT ON THE RESULTS.}

\begin{figure}[!ht]
    \centering
    \includegraphics[width=0.9\linewidth]{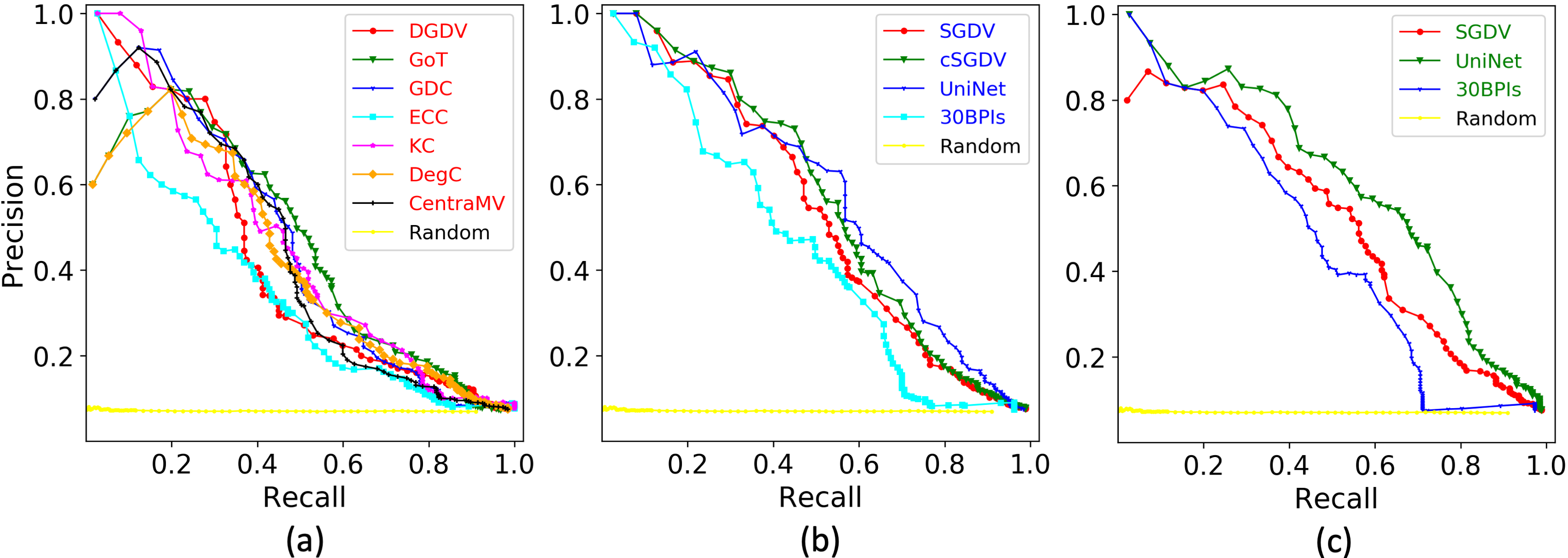}
    \vspace{-0.2cm}
    \caption{\textbf{The precision-recall curves} of all considered features (each under the best dimensionality choice and classifier) when run on their respective networks, plus the random approach, when using \emph{GenAge} to define aging- and non-aging-related genes, and when reducing feature dimensionality \emph{during} the 5-fold cross-validation. Panel (a) is for the seven dynamic features when run on the dynamic aging-specific subnetwork. Panel (b) is for the four static features when run on the static aging-specific subnetwork. Panel (c) is for the three of the four static features (all but cGDV) that could be run on the entire static context-unspecific network. In each panel, the result for the random approach is also shown. On a given curve, a dot corresponds to a prediction threshold $g$, where we vary $g$ in increment of 2 from 1 to 60, and in increment of 10 from 60 to 500. So, the dots from left to right correspond to $g$ from 1 to 500. Analogous results when using \emph{GTEx-DAG} to define aging- and non-aging-related genes or when reducing feature dimensionality \emph{prior to} the 5-fold cross-validation are shown in Supplementary Figures S3, S10,  and S13. } 
    \label{fig:GenAge-PRC}
    \end{figure}

Henceforth, we analyze each feature under its best dimensionality\_choice--classifier combination (that maximizes AUPR). The corresponding precision-recall curve that has led to selecting this combination  is shown in Figure \ref{fig:GenAge-PRC}.

\begin{table}[!htp]
    \caption{The selected (best) dimensionality choice  and classifier combination for each feature. In ``$X + Y (Z)$'', $X$ is the best dimensionality choice, $Y$ is the best classifier, and $Z$ is the resulting dimension (in the case of PCA, the dimension can vary between the five folds of the cross-validation and so $Z$ is the resulting dimension averaged over all five folds). ``None'' means that performing no dimensionality reduction was better than both  PCA and tSNE. Red-colored features are dynamic features extracted from the dynamic aging-specific subnetwork. Blue-colored features are static features extracted from the static aging-specific subnetwork. Green-colored features are static features extracted from the entire context-unspecific network. Underlined features are existing features (the others are ours).}
    \vspace{-0.1cm}
    \label{table:classifiers}
    \scriptsize
    \centering
        \begin{tabular}{ccccc}
        \hline
        \textcolor{red}{DGDV} & \textcolor{red}{GoT} & \textcolor{red}{GDC} & \textcolor{red}{ECC} & \textcolor{red}{KC}  \\ 
        None + AB (3,727) & PCA + LR (2) & None + NB (37) & None + RF (37) & None + RF (37) \\ \hline
        \textcolor{red}{DegC} & \textcolor{red}{CentraMV} & \textcolor{blue}{SGDV} & \textcolor{blue}{cSGDV} & \textcolor{blue}{\underline{UniNet}}  \\ 
         PCA + LR (1.2) & PCA + MLP (1) & PCA + LR (2) & PCA + LR (2) & None + AB (18)  \\ \hline
        \textcolor{blue}{\underline{30BPIs}}     &  {\color[HTML]{19901A} SGDV} & {\color[HTML]{19901A} \underline{UniNet}} & {\color[HTML]{19901A} \underline{30BPIs}} \\
        None + LR (30) & PCA + LR (1.4)   & None + AB (18) & PCA + LR (24) \\ \hline
    \scriptsize
    \end{tabular}
    \vspace{-0.05in}
\end{table}

%\vspace{-0.1cm}

\subsection{Using an aging-specific subnetwork vs. using the entire context-unspecific network \label{subsect:stat-vs-entire}}

For all three hypotheses, including hypothesis 1 that is tested in this section, we compare the performance of the 14 features (each under its best dimensionality and classifier choice), as well as of the random approach, as follows. First, we perform evaluation via the 5-fold cross validation, measuring AUPR over all prediction thresholds, as well as precision, recall, and F-score at the prediction threshold where F-score is maximized. Also, we perform validation of novel predictions on the cancer data, again measuring precision, recall, and F-score (AUPR is not applicable to the cancer-related validation, as this validation deals only with the single prediction threshold where the 5-fold cross-validation F-score is maximized).

To validate hypothesis 1, it would suffice for the best (our or existing) feature on the dynamic or static aging-specific subnetwork to outperform every feature on the entire context-unspecific network. We find that hypothesis 1 holds in both the 5-fold cross-validation and the cancer-related validation, with respect to almost all of the measures, as follows.

\begin{figure}[!ht]
    \centering
    \includegraphics[width=0.75\linewidth]{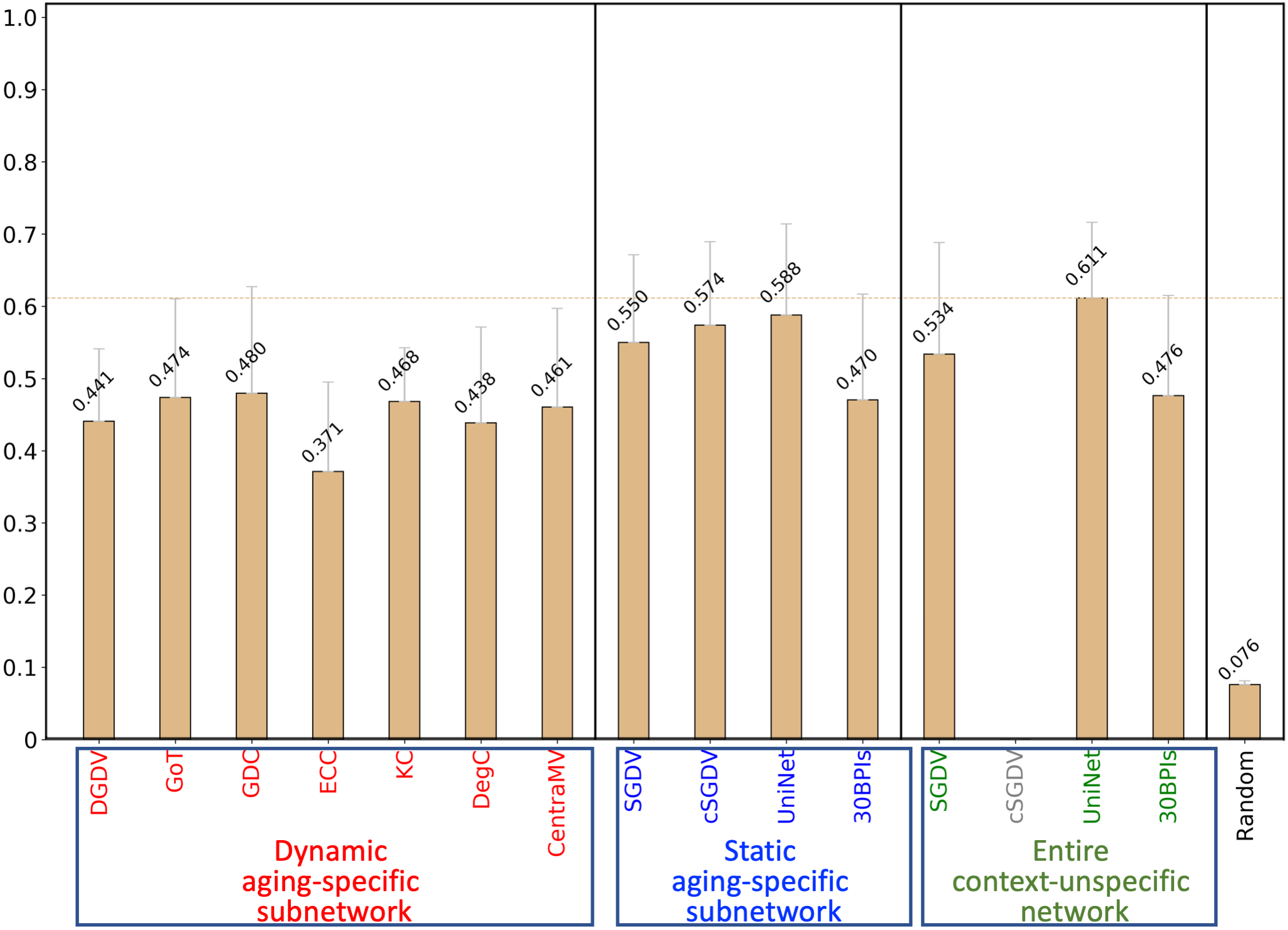}
    %\vspace{-0.4cm}
    \caption{\textbf{Prediction accuracy in the 5-fold cross-validation in terms of AUPR} of all considered features (each under the best dimensionality choice and classifier) when run on  their respective networks, plus the random approach, when using \emph{GenAge} to define aging- and non-aging-related genes, and when reducing feature dimensionality \emph{during} the 5-fold cross-validation. The seven dynamic features can be run only on the dynamic subnetwork. The four static features can be run on  the static subnetwork as well as the entire network. The result for the gray-colored cSGDV on the entire network is missing because this network is too large to run cSGDV on it. For convenience, we show the AUPR values on top of the bars. Analogous results when using \emph{GTEx-DAG} to define aging- and non-aging-related genes or when reducing feature dimensionality \emph{prior to} the 5-fold cross-validation are shown in Supplementary Figures S4, S11, and S14. }
    \label{fig:aupr-main}
    \end{figure}

\begin{figure}[!ht]
    \centering
    \includegraphics[width=0.75\linewidth]{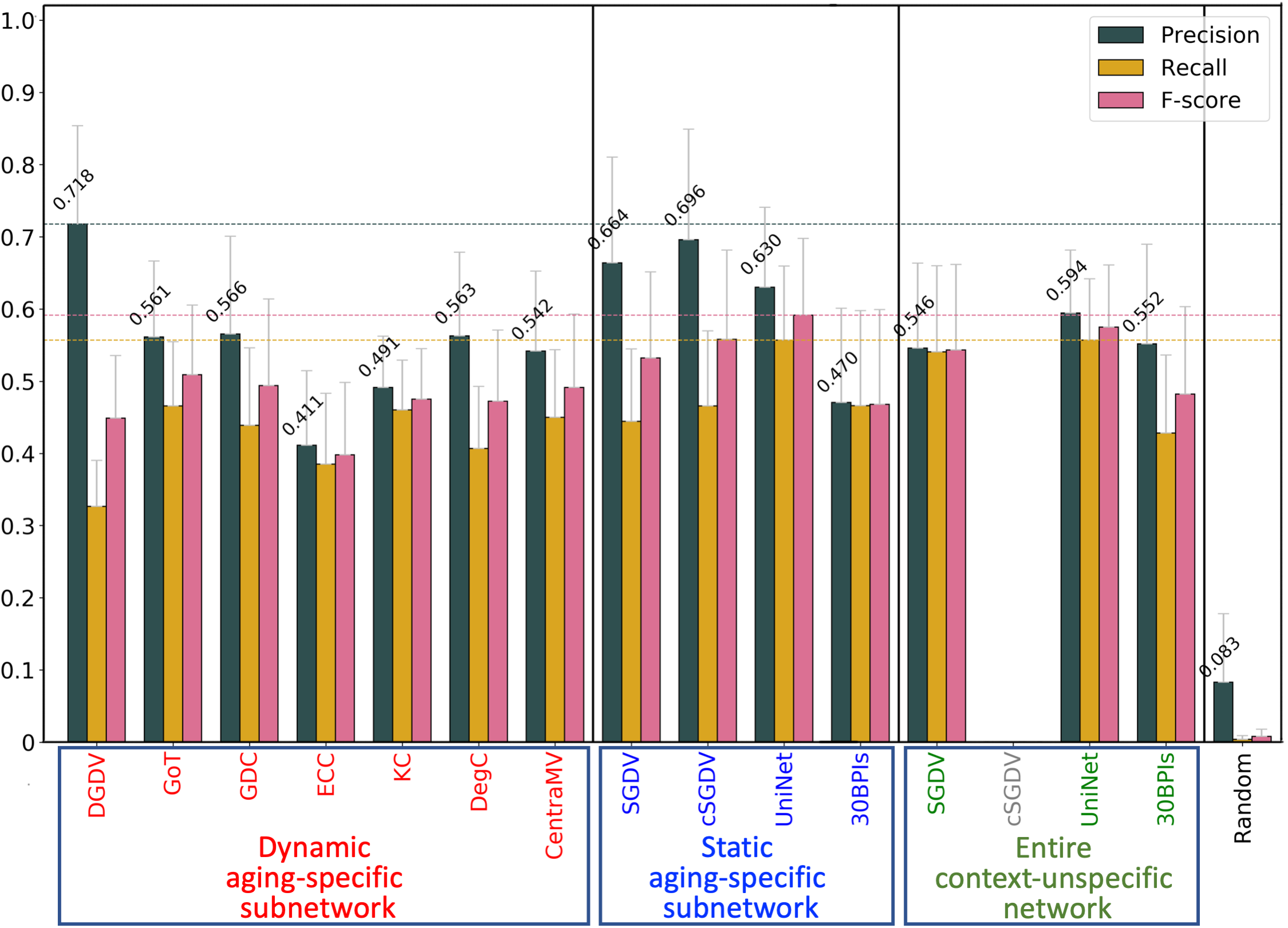}
  %  \vspace{-0.4cm}
    \caption{\textbf{Prediction accuracy  in the 5-fold cross-validation in terms of precision, recall, and F-score} of all considered features (each under the best dimensionality choice and classifier)  when run on  their respective networks, plus the random approach, when using \emph{GenAge} to define aging- and non-aging-related genes, and when reducing feature dimensionality \emph{during} the 5-fold cross-validation. The seven dynamic features can be run only on the dynamic subnetwork. The four static features can be run on  the static subnetwork as well as the entire network. The result for the gray-colored cSGDV on the entire  network is missing because this network is too large to run cSGDV on it. For convenience, we show the precision values on top of the corresponding bars. Analogous results when using \emph{GTEx-DAG} to define aging- and non-aging-related genes or when reducing feature dimensionality \emph{prior to} the 5-fold cross-validation are shown in Supplementary Figures S5, S12, and S15.  }
    \label{fig:maxF1-main}
    \end{figure}

\underline{In the 5-fold cross-validation}, in terms of AUPR, it is UniNet on the entire context-unspecific network that performs the best (Figure \ref{fig:aupr-main}). But its performance is tied to those of UniNet, cSGDV, and SGDV  on the static aging-specific subnetwork, meaning that UniNet's observed superiority when run on the entire context-unspecific network is \emph{not} statistically significant. Nonetheless, hypothesis 1 does not hold in terms of AUPR. In terms of precision, recall, and F-score, there exists a feature on an aging-specific subnetwork (namely UniNet on the static subnetwork) that is at least as good or (marginally, though not statistically significantly) better than every feature on the entire context-unspecific network in terms of \emph{all three measures simultaneously} (Figure \ref{fig:maxF1-main}). Hence, hypothesis 1 holds in terms of all of precision, recall, and F-score as a whole. Furthermore, when looking at precision \emph{alone}, which we favor over recall (and thus over F-score as well as AUPR) for reasons discussed below, DGDV on the dynamic aging-specific subnetwork, as well as cSGDV, SGDV, and UniNet on the static aging-specific subnetwork, perform better than all features on the entire context-unspecific network. Of these four features, DGDV, SGDV, and cSGDV (though not UniNet) on their respective aging-specific subnetworks significantly outperform SGDV and 30BPIs on the entire context-unspecific network (adjusted $p$-values below 0.039), although they marginally outperform UniNet on the entire context-unspecific network. So, hypothesis 1 holds in terms of precision.

Recall that AUPR summarizes the performance of a dimensionality\_choice--classifier combination over the entire  $g$ (prediction threshold) range (Figure \ref{fig:GenAge-PRC}). That is, by relying on AUPR, one cannot produce a specific predicted gene set for further analysis. In our study, we use AUPR mainly as a parameter to help us select the best dimensionality\_choice--classifier combination for each feature. Then, for a given feature's selected combination, our goal is to choose a good prediction threshold $g$ in order to identify confident novel aging-related genes for future wet lab validation. To do this, i.e., to produce a prediction set for a given feature, we choose $g$ where F-score is maximized. Because  F-score balances between precision and recall, this ensures that at our selected $g$ for a given feature, neither precision or recall will be low. But once such a $g$ value is selected, we believe that precision should be favored over recall (and thus also over F-score and AUPR, both of which rely on recall in addition to precision). This is because we believe that in biomedicine, for wet lab validation of predictions, it is more important to have fewer (though still sufficiently many!) but more correct predictions than many but largely incorrect predictions. For example, one can predict all possible genes as aging-related and thus yield  maximum recall of 100\%, but at the same time, this would result in randomly low precision. Given that in our evaluation, all features have reasonably high recall, we believe that precision should be the deciding factor of which feature is the best.

\begin{figure}[!ht]
    \centering
    \includegraphics[width=0.75\linewidth]{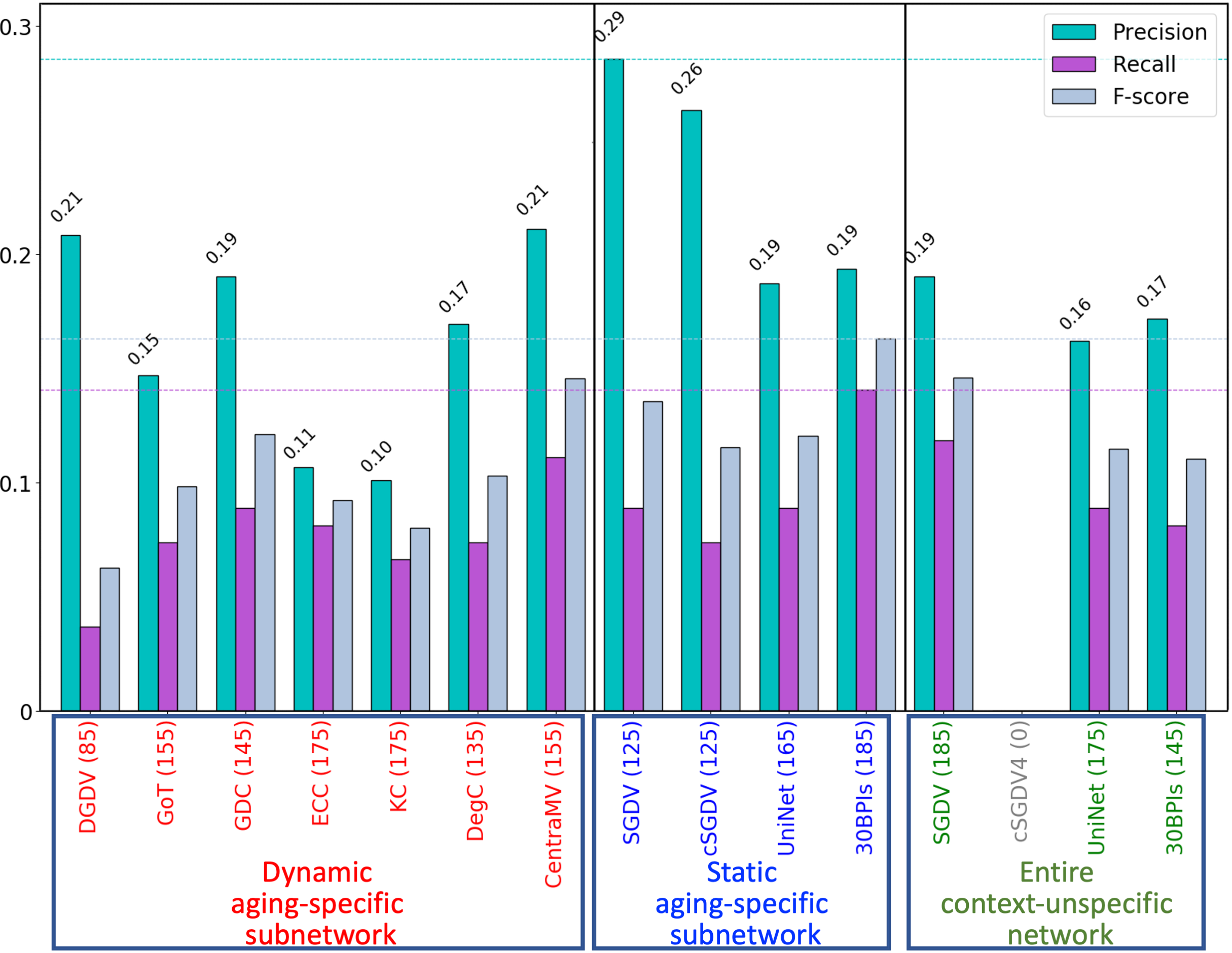}
    %\vspace{-0.4cm}
    \caption{
    \textbf{Prediction accuracy in the validation of newly predicted genes via \emph{all driver} cancer genes, in terms of precision, recall, and F-score} of all considered features (each under the best dimensionality choice and classifier) when run on their respective networks, when using \emph{GenAge} to define aging- and non-aging-related genes, and when reducing feature dimensionality \emph{during} the 5-fold cross-validation. The seven dynamic features can be run only on the dynamic subnetwork. The four static features can be run on  the static subnetwork as well as the entire network. The result for the gray-colored cSGDV on the entire  network is missing because this network is too large to run cSGDV on it. On the $x$-axis, a given feature's name is followed by the number of the feature's newly predicted genes across all five cross-validation folds combined (in parentheses). For convenience, we show the precision values on top of the corresponding bars. Analogous results for \emph{mutation driver} (rather than all driver) cancer genes or when reducing feature dimensionality \emph{prior to} the 5-fold cross-validation are shown in Supplementary Figures S1, S6, and S7. }
    \label{fig:cancers}
    \end{figure}

\underline{In the cancer-related validation}, as shown in Figure \ref{fig:cancers}, in terms of precision, SGDV on the static aging-specific subnetwork is by far the best of all features, followed by cSGDV on the same subnetwork, followed by DGDV and CentraMV on the dynamic subnetwork; and all four of these features on their aging-specific subnetworks are superior to all features on the entire context-unspecific network. Hence, hypothesis 1 also holds in terms of the cancer-related validation with respect to precision. Also, while we favor precision over recall (and thus over F-score), hypothesis 1 also holds in the cancer-related validation with respect to precision, recall, and F-score \emph{as a whole}. That is, there exists a feature on an aging-specific subnetwork (namely 30BPIs on the static subnetwork) that is at least as good or better than every feature on the entire context-unspecific  networks in terms of \emph{all three measures simultaneously} (Figure \ref{fig:cancers}). Note that these results are for the case of using \emph{all driver genes} in the cancer-related validation (Figure \ref{fig:cancers}). Results are qualitatively similar in the case of using \emph{mutation driver genes} (Supplementary Figure S1).

\subsection{Using our features vs. using the existing features}

To validate hypothesis 2, it would suffice for one or more of our nine features on the dynamic or static aging-specific subnetwork to be superior to both of the existing features on the static aging-specific subnetwork.
Indeed, we find this to be the case in both the 5-fold cross-validation and the cancer-related validation, with respect to at least one of the measures, as follows.

\underline{In the 5-fold cross-validation}, in terms of precision, three of our features (DGDV, SGDV, and cSGDV) outperform both UniNet and 30BPIs (Figure \ref{fig:maxF1-main}). Their superiority is statistically significant over 30BPIs (adjusted $p$-values below 0.039) and marginal over UniNet. So, hypothesis 2 holds in terms of precision, although not in terms of recall, F-score, and AUPR (Figures \ref{fig:aupr-main} and \ref{fig:maxF1-main}).

\underline{In the cancer-related validation} (Figure \ref{fig:cancers}), in terms of precision, SGDV and cSGDV on the static aging-specific subnetwork as well as DGDV and CentraMV on the dynamic subnetwork are superior to both UniNet and 30BPIs on the static subnetwork. So, hypothesis 2 holds in terms of precision,  although not  in terms of recall and F-score (Figure \ref{fig:cancers}). Note that these results are for the case of using all driver genes in the cancer-related validation (Figure \ref{fig:cancers}). Results are qualitatively similar for the case of using mutation driver genes (Supplementary Figure S1).

Hence, hypothesis 2 holds in both the evaluation/validation tests in terms of precision, which we again remind is the measure that we believe should be favored over recall (and thus over F-score and AUPR).

Importantly, all features perform significantly better than at random (adjusted $p$-values below 0.039), while at the same time, each of them is far from perfect (Figure \ref{fig:maxF1-main}). Moreover, when we compute overlaps between the different features' true positive gene predictions (i.e., actual aging-related genes present in GenAge), we find that while all pairwise overlaps are statistically significantly high (adjusted $p$-values $ \leq 3.8 \times 10^{-6}$), their sizes are between 45\% and 95\%, depending on the compared features; the average overlap size over all feature pairs is 63\%  (Supplementary Figure S2). Thus, the different features' predictions are relatively complementary to each other even in terms of their true positive predictions. The features are even more complementary in terms of their newly predicted aging-related genes (that are not currently linked to aging in any of the six considered ground truth data sets): their pairwise overlaps are between 7\% and 81\%, with the average overlap size over all feature pairs of 24\% (Supplementary Figure S2). These results indicate that the features might be capturing somewhat complementary aspects of network topology. So, developing a new (e.g., ensemble learning) approach that would combine their complementary topological aspects might be promising.

\begin{figure}[!ht]
    \centering
    \includegraphics[width=0.4\linewidth]{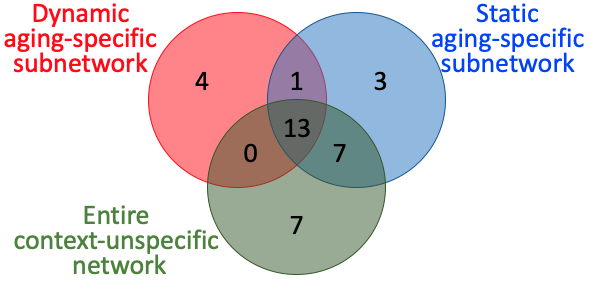}
    \caption{\textbf{Overlaps of the cancer data-validated newly predicted genes resulting from the three networks}, when using \emph{GenAge} to define aging- and non-aging-related genes, and when reducing feature dimensionality \emph{during} the 5-fold cross-validation. Each of the three prediction sets includes (novel aging-related and cancer-validated) genes that are predicted from the corresponding network by \emph{any} applicable feature. For example, the $4+1+0+13=18$  genes under ``dynamic aging-specific subnetwork'' correspond to the union over all seven dynamic features of all of their newly predicted genes that are validated in the cancer data.   Analogous results when reducing feature dimensionality \emph{prior to} the 5-fold cross-validation are shown in Supplementary Figure S8.} 
    \label{fig:venn}
\end{figure}

Because the different features are complementary, i.e.,  because the features run on any one of the three networks predict novel aging-related genes that the features run on the other two networks do not (Figure \ref{fig:venn}), we combine all of their newly predicted genes. Of these, we consider those 35 genes that are present in (i.e., validated by) the cancer data to be good targets for future wet lab validation by the community (Table \ref{table:genes}).

\begin{table}[!htp]
\centering
\caption{The 35 newly predicted genes (i.e., novel aging-related predictions) across all considered features that are also cancer driver genes. Of these, \emph{mutation} cancer driver genes are bolded. In parentheses are the number of features (out of 14 possible features) that support a prediction, and up to three stars, where the presence of a red star indicates that a given gene is predicted from the dynamic aging-specific subnetwork, the presence of a blue star indicates that  a given gene is predicted from the static aging-specific subnetwork, and the presence of a green star indicates that  a given gene is predicted from the entire context-unspecific network; a missing star indicates that a given genes is not predicted from the corresponding network. The genes are listed in alphabetical order (left to right, then top to bottom).} 
\vspace{-0.1cm}
\label{table:genes}
\scriptsize
\begin{tabular}{|r|r|r|r|r|r|r|}
\hline
 ATF1 (1\textcolor{red}{$*$})& BIRC6 (2\textcolor{red}{$*$}\textcolor{blue}{$*$}) & BTK (3\textcolor{blue}{$*$}{\color[HTML]{19901A}$*$}) & CASP9 (9\textcolor{red}{$*$}\textcolor{blue}{$*$}{\color[HTML]{19901A}$*$}) & \textbf{CCND1} (12\textcolor{red}{$*$}\textcolor{blue}{$*$}{\color[HTML]{19901A} $*$}) \\ \hline 
 
  \textbf{CDH1} (5\textcolor{blue}{$*$}{\color[HTML]{19901A} $*$}) &
 \textbf{DAXX} (13\textcolor{red}{$*$}\textcolor{blue}{$*$}{\color[HTML]{19901A} $*$}) & DDX5 (2{\color[HTML]{19901A} $*$})& EZR (6\textcolor{red}{$*$}\textcolor{blue}{$*$}{\color[HTML]{19901A} $*$}) & FBXO11 (2{\color[HTML]{19901A} $*$})  \\ \hline 
 
 HSP90AB1 (5\textcolor{red}{$*$}\textcolor{blue}{$*$}{\color[HTML]{19901A} $*$}) & \textbf{IKZF1} (1\textcolor{blue}{$*$}) & IRS4 (7\textcolor{red}{$*$}\textcolor{blue}{$*$}{\color[HTML]{19901A} $*$}) & KDR (11\textcolor{red}{$*$}\textcolor{blue}{$*$}{\color[HTML]{19901A} $*$}) & \textbf{KLF4} (1{\color[HTML]{19901A} $*$}) \\ \hline 
 
 LEF1 (4\textcolor{blue}{$*$}{\color[HTML]{19901A} $*$}) & \textbf{MEN1} (1\textcolor{blue}{$*$}) & MGMT (2{\color[HTML]{19901A} $*$}) & 
  MUC1 (2\textcolor{blue}{$*$}{\color[HTML]{19901A} $*$}) & MYST2 (3\textcolor{red}{$*$}) \\ \hline
  
 \textbf{NCOA3} (12\textcolor{red}{$*$}\textcolor{blue}{$*$}{\color[HTML]{19901A} $*$}) & NFKBIE (1\textcolor{red}{$*$})& PABPC1 (6\textcolor{red}{$*$}\textcolor{blue}{$*$}{\color[HTML]{19901A} $*$}) & REL (4\textcolor{red}{$*$}\textcolor{blue}{$*$}{\color[HTML]{19901A} $*$}) & 
 RHOA (1{\color[HTML]{19901A} $*$})  \\ \hline
 
 SFRS3 (1\textcolor{blue}{$*$}) & SGK1 (2\textcolor{blue}{$*$}{\color[HTML]{19901A} $*$}) & \textbf{SMAD2} (14\textcolor{red}{$*$}\textcolor{blue}{$*$}{\color[HTML]{19901A} $*$}) & \textbf{SMAD4} (13\textcolor{red}{$*$}\textcolor{blue}{$*$}{\color[HTML]{19901A} $*$}) & SMARCD1 (1{\color[HTML]{19901A} $*$}) \\ \hline
 
 TRIM24 (3\textcolor{red}{$*$}\textcolor{blue}{$*$}{\color[HTML]{19901A} $*$}) & TRIM27 (1{\color[HTML]{19901A} $*$}) & \textbf{TSC1} (1\textcolor{red}{$*$}) & VAV1 (7\textcolor{blue}{$*$}{\color[HTML]{19901A} $*$}) & WAS (5\textcolor{blue}{$*$}{\color[HTML]{19901A} $*$})  \\ \hline
 
\end{tabular}
\scriptsize
%\vspace{-0.05in}
\end{table}
    
\subsection{Using the dynamic aging-specific subnetwork vs.using the static aging-specific subnetwork}\label{sect:results_dynamic_static}

To validate hypothesis 3, it would suffice for the best feature on the dynamic subnetwork to be superior to the best feature on the static subnetwork. We find this to hold in the 5-fold cross validation in terms of one of the measures. However, this does not hold in the cancer-related validation with respect to any measure.

\underline{In the 5-fold cross-validation}, in terms of precision, DGDV on the dynamic subnetwork outperforms all features on the static subnetwork (Figure \ref{fig:maxF1-main}), although its superiority is not statistically significant. So, hypothesis 3 holds in terms of precision, although not in terms of recall, F-score, and AUPR (Figures \ref{fig:aupr-main} and \ref{fig:maxF1-main}).

\underline{In the cancer-related validation}, SGDV and cSGDV on the static subnetwork have higher precision  than all features on the dynamic subnetwork, and 30BPIs on the static subnetwork has higher recall and F-score  than all features on the dynamic subnetwork (Figure \ref{fig:cancers}). Hence, hypothesis 3 does not hold in the cancer-related validation. Note that these results are for the case of using all driver genes in the cancer-related validation (Figure \ref{fig:cancers}). Results are qualitatively similar in the case of using mutation driver genes (Supplementary Figure S1).

Even though some of the dynamic features outperform all  static features with respect to only precision but not the other measures, and only in the 5-fold cross-validation but not the cancer-related validation,   the dynamic features' predictions complement those of the static features, i.e., they have a potential to identify novel aging-related genes that the static features might miss. In fact, all seven dynamic features combined do predict seven true positives, i.e., \emph{known} aging-related genes, as well as four novel aging-related genes that can validated in the cancer data, that the static features miss.

%\textcolor{orange}{However, we find this not to be the case. The static SGDV has a higher precision in the 5-fold cross validation (Figure \ref{fig:maxF1-main}) and a higher cancer-data validation rate (Figure \ref{fig:cancers}) than any dynamic feature. Even though our  dynamic features do not outperform SGDV, we remind that the dynamic features' predictions complement those of SGDV, i.e., they have a potential to identify novel aging-related genes that the static feature might miss. In fact, the dynamic features combined do predict 17 true positives (i.e, \emph{known} aging-related genes) that SGDV misses.}

The fact that the best feature on the dynamic aging-specific subnetwork is not consistently superior to the best feature on the static aging-specific subnetwork is surprising, as the former should be capturing more of the aging-related information, given that the aging process is dynamic and not static. Examining the cause(s) of this surprising result is important. This is the subject of future work. A reason for it could be that the current dynamic aging-specific subnetwork is suboptimal, as it is constructed using the induced approach \cite{faisal2014dynamic}, meaning that \emph{all} edges between \emph{only} those genes that are significantly expressed at a given age are considered. More recently, the notion of network propagation was introduced as a better alternative, but for inference of only a \emph{static} context-specific subnetwork, and mainly in the context of \emph{cancer} \cite{cowen2017network}. Network propagation diffuses gene activities (expression levels) over the entire network to assign context-specific weights to its edges, and then it uses only the highest-weight edges, possibly between some non-significantly expressed genes in addition to significantly expressed ones, to form a static context-specific subnetwork. Because current network propagation cannot infer a dynamic subnetwork, we cannot consider it in this study. In fact, developing a novel network propagation approach for inference of a dynamic aging-specific subnetwork is the subject of our on-going work \cite{newaz2018improving} that is orthogonal to this study and is consequently well beyond the scope and timeline of this paper. Note that considering a static network propagation-based aging-specific subnetwork in this study is unnecessary, because it could not help with testing our hypothesis 3. Instead, it could ``only'' further strengthen our hypotheses 1 and 2, which we have already proven in many/most of the evaluation tests. And it is proving hypothesis 1 or hypothesis 2 that is necessary and sufficient to justify the need for introducing our AGENT framework.

\subsection{The effect on results of reducing feature dimensionality prior to vs. during the 5-fold cross-validation}\label{sect:pretrainreduction}

Due to the open debate regarding whether it is more appropriate to reduce feature dimensionality prior to vs. during the 5-fold cross-validation, we consider both options. All results reported thus far in the paper have been for the latter. In this section, we report results for the former. 

We find that when reducing feature dimensionality prior to the 5-fold cross-validation (Supplementary Figures S4, S5, and S6), almost all results remain both qualitatively and quantitatively the same as when reducing feature dimensionality during the 5-fold cross-validation (Figures \ref{fig:aupr-main},  \ref{fig:maxF1-main}, and \ref{fig:cancers}). That is, the conclusions about all three hypotheses remain the same.  

Note that it is actually surprising that reducing feature dimensionality prior to the 5-fold cross-validation does not perform better than doing so during the 5-fold cross-validation. This is because the former may leak some of the testing information into the training process, while the latter does not. Yet, the latter is as accurate as the former.

\subsection{The effect on results of using the secondary GTEx-DAG-based definition of aging- and non-aging-related genes}\label{sect:secondaryGTEx-result}

All of the results reported thus far have been with respect to the aging-related ground truth knowledge from GenAge. Because using this knowledge that originates from model species to predict aging-related human genes might miss some human-unique aspects of the aging process, in this section, we consider the secondary aging-related ground truth data that was obtained directly in human, namely GTEx-DAG. When using the GTEx-DAG-based definition of aging- vs. non-aging-related genes, we compare the different features in terms of the 5-fold cross-validation but not the cancer-related validation, for the reasons discussed below. 

In the 5-fold cross-validation (Supplementary Figures S10--S15), at the first, superficial look, it seems that hypothesis 1 holds (marginally though not statistically significantly) with respect to all measures (precision alone, all of precision, recall, and F-score as a whole, and AUPR). Also, hypotheses 2 and 3 hold (marginally though not statistically significantly) in terms of AUPR but not in terms of any of precision, recall, or F-score. These results hold when reducing feature dimensionality during (Supplementary Figures S11--S12) as well as prior to (Supplementary Figures S14--S15)   the 5-fold cross-validation. 

However, at the second, closer look, we shockingly find that when using the GTEx-DAG data, the prediction accuracy for all features is almost always under 30\% (compared to 60\%-70\% accuracy when using the GenAge data). That is, when using the GTEx-DAG data, none of the features is significantly better than the random approach in terms of any of precision (adjusted $p$-values $\geq 0.071$), recall (adjusted $p$-values $\geq 0.069$), F-score (adjusted $p$-values $\geq 0.054$), or AUPR (adjusted $p$-values $\geq 0.071$). This holds for when reducing feature dimensionality during as well as prior to the 5-fold cross-validation (Supplementary Figures S10--S15). So, actually, upon this closer look, it makes no sense to compare the different features in order to determine whether any one of the three hypotheses hold.

Because the performance is random-like for all features when using the GTEx-DAG data to train and test the classifiers, we do not  make any novel predictions from the GTEx-DAG-based classifiers. Consequently, the cancer-data validation tests cannot be performed.

This result is shocking to us. Our intuition says that using GTEx-DAG, aging-related ground truth data obtained \emph{directly in human} (via differential gene expression analyses), as node labels \emph{in the human} PPI network to predict \emph{human} aging-related genes should result in higher accuracy than at random and even in higher accuracy than using GenAge, aging-related ground truth data obtained \emph{in model species and then transferred to human} (via genomic sequence alignment). Also, our intuition says that using GTEx-DAG, a \emph{gene expression-based} aging-related data set, to predict from the aging-related subnetworks that have been obtained by combining PPI data with (different, independent) aging-related \emph{gene expression} data, should yield higher accuracy than using the \emph{sequence-based} GenAge. But our results contradict our 
intuition. We could think of two reasons that could be behind this shocking result. 

First, it could be that GenAge is a higher-confidence aging-related ground truth data set than GTEx-DAG, i.e., that GenAge is more likely to contain true aging-related information, possibly because the sequence-based analyses are more reliable than  the expression-based analyses. Indeed, GenAge is considered by many to be one of the highest-confidence aging-related ground truth data sets, if not the highest-confidence one, which is why we have
primarily focused on this data in the first place. 

Second, note that GenAge and GTEx-DAG are quite complementary. Of the 187 aging-related genes from GenAge genes and the 439 aging-related genes from GTEx-DAG genes, only 16 are in their overlap. So, it could be that both GenAge and GTEx-DAG are capturing true aging-related information well, but because they are complementary, it could be that the topology of the networks we considered is ``encoding'' better the aging-related genes from GenAge than those from GTEx-DAG. In turn, this would allow network features to better ``recognize'' and thus more correctly classify the aging-related genes from GenAge.  But what could cause this in the first place? We offer two potential explanations.

The first potential explanation is as follows.  The  aging-specific subnetworks that we use, which were inferred in a past study from even an older human PPI network (from HPRD), are quite old. Use of biotechologies for sequence data collection and computational approaches for sequence data analyses may have been more popular in the past than use of those for expression data collection and analyses. Because of these two factors, it is possible that biotechological discovery of PPIs in the quite old PPI network data was guided more by output of sequence data analyses than by output of expression data analyses. If this is the case, then features extracted from such a  sequence-biased PPI network would be expected to match better sequence-based node labels (i.e., aging-related knowledge from GenAge) than expression-based labels. Newer human PPI network data have become available \cite{das2012hint,rolland2014proteome}, including a fresh-off-the-press reference map of the human binary (i.e., yeast two-hybrid) PPI network  \cite{luck2020reference}. Discovery of such data has been guided by newer, state-of-the-art biotechologies or data analyses. Consequently, such data is likely less noisy, i.e., of higher confidence, as well as possibly better ``aligned'' with the recent GTEx-DAG gene expression-based aging-related ground truth knowledge. So, using such newer data to infer new aging-specific PPI subnetworks  might yield higher prediction accuracy when the GTEx-DAG data is used to make aging-related predictions from the new subnetworks. Testing this would certainly be important. However, inference of a new aging-specific subnetwork is not the focus of this study and is thus out of the scope of this paper. Instead, that topic is left for future work. 

The second potential explanation is as follows. The  aging-related gene expression data used to construct the aging-specific subnetworks we have considered was obtained via a quite old biotechology, namely microarrays \cite{berchtold2008gene}. On the other hand, the GTEx-DAG aging-related ground truth expression data was obtained via a different, newer and more reliable biotechology, namely RNA-seq \cite{jia2018analysis}. The two techologies might be covering different functional slices of the aging process in the cell, with at least one of them containing noise, i.e., false positives as well as false negatives. So, there might be a (large) mismatch between the aging-related knowledge embedded into the topology of the considered PPI subnetworks and the aging-related knowledge used as labels on nodes of the networks for the classification purpose. However, while this mismatch could explain why we get poor results when using GTEx-DAG, it could still  not explain why using the \emph{sequence-based} GenAge data yields much higher-quality predictions than using \emph{any} gene expression-based data, including GTEx-DAG.

%\textcolor{blue}{HERE, DISCUSS RESULTS FOR GTEX POST-TRAIN AND THEN FOR GTEX PRE-TRAIN}

%%%%%%%%%%%%%%%%%%%%%%%%%%%%%%%%%%%%%%%%%%%%%%%%%%%%%%%%%%%%%%%%%%%%%%%%%%%%%%%%%%%%%%%%%%%%%%%%%%%%%%%%%%%%%%%%%%%%%%%%%%%%%%%%%%%%%%%%%%%%%%%%%%%%%%%%%%%%%%%%%%%%%%%%%%%%%%
\section{Conclusions}
In this paper, we propose a new supervised  framework called AGENT that utilizes node features extracted from an aging-specific PPI subnetwork in the context of classification  to predict human aging-related genes. In a fair and comprehensive evaluation, we show that in many of the evaluation tests, our framework results in higher prediction accuracy than existing state-of-the-art methods for the same purpose that use an entire static context-unspecific PPI network (i.e., hypotheses 1 and 2 hold overall). Our comprehensive evaluation has opened up several interesting research questions for future examination, as follows. A newer entire context-unspecific PPI network or newer aging-related gene expression data can be used to infer newer aging-specific subnetworks. A novel (e.g., network propagation-based) approach for inference of a dynamic aging-specific subnetwork can be developed (rather than using the only current option - the induced approach). These two future directions could help with hypothesis 3, i.e., they may yield a dynamic aging-related subnetwork that would be superior to all static aging-related subnetworks, as well as explain why using GTEx-DAG, the recent expression-based aging-related ground truth data set, has yielded random-like prediction accuracy from the current aging-specific networks. Also, given a relatively high complementary of the aging-related predictions produced the different features, it might be worth to pursue development of a novel feature or an ensemble learning approach that would integrate the complementary aspects of the existing features. Moreover, while in this paper we  focus on \emph{human} aging, our work can  easily  be applied to other species. Further, it can be applied to dynamic biological processes other than aging, such as disease progression. Also, while we  focus on \emph{PPI} networks, our work can  easily  be applied to other types of biological networks.

\section*{Competing interests}
  The authors declare that they have no competing interests.

\section*{Author's contributions}
QL and TM conceived and designed the study. QL performed all computational experiments and produced all results. QL and TM analyzed the results and their implications. QL and TM wrote the paper. TM supervised the study. 

\section*{Acknowledgements}
This work is supported by the U.S. National Science Foundation (NSF) CAREER award (CCF-1452795).

\bibliographystyle{unsrt}  
\bibliography{references}

%%%%%%%%

\end{document}